\documentstyle[epsfig]{aipproc}
\begin{document}
\input psfig.tex
\title{Evidence from Type Ia Supernovae for an Accelerating Universe}

\author{Alexei V. Filippenko$^*$}
\address{Department of Astronomy, University of California,
Berkeley, CA 94720-3411 (alex@astro.berkeley.edu)}
\author{Adam G. Riess$^*$}
\address{Space Telescope Science Institute, 3700 San Martin Dr., Baltimore, 
MD 21218 (ariess@stsci.edu)}
\address{
($^*$On behalf of the High-$z$ Supernova Search Team)}

\maketitle

\begin{abstract} We review the use of Type Ia supernovae for cosmological
distance determinations.  Low-redshift SNe~Ia ($z \lesssim 0.1$) demonstrate
that the Hubble expansion is linear, that $H_0 = 65 \pm 2$ (statistical) km
s$^{-1}$ Mpc$^{-1}$, and that the properties of dust in other galaxies are
similar to those of dust in the Milky Way.  We find that the light curves of
high-redshift ($z = 0.3$--1) SNe~Ia are stretched in a manner consistent with
the expansion of space; similarly, their spectra exhibit slower temporal
evolution (by a factor of $1 + z$) than those of nearby SNe~Ia.  The luminosity
distances of our first set of 16 high-redshift SNe~Ia are, on average, 10--15\%
farther than expected in a low mass-density ($\Omega_M=0.2$) universe without a
cosmological constant.  Preliminary analysis of our second set of 9 SNe~Ia is
consistent with this. Our work supports models with positive cosmological
constant and a current acceleration of the expansion.  We address the main
potential sources of systematic error; at present, none of them appears to
reconcile the data with $\Omega_\Lambda=0$ and $q_0 \geq 0$.  The dynamical age
of the Universe is estimated to be $14.2 \pm 1.7$ Gyr, consistent with the ages
of globular star clusters.

\end{abstract}

\section*{ INTRODUCTION}

    Supernovae (SNe) come in two main varieties (see reference \cite{avf97b} for
a review). Those whose optical spectra exhibit hydrogen are classified as Type
II, while hydrogen-deficient SNe are designated Type I. SNe~I are further
subdivided according to the appearance of the early-time spectrum: SNe~Ia are
characterized by strong absorption near 6150~\AA\ (now attributed to Si~II),
SNe~Ib lack this feature but instead show prominent He~I lines, and SNe~Ic have
neither the Si~II nor the He~I lines. SNe~Ia are believed to result from the
thermonuclear disruption of carbon-oxygen white dwarfs, while SNe~II come from
core collapse in massive supergiant stars. The latter mechanism probably
produces most SNe~Ib/Ic as well, but the progenitor stars previously lost their
outer layers of hydrogen or even helium.

   It has long been recognized that SNe~Ia may be very useful distance
indicators for a number of reasons; see \cite{bra92,bra98},
and references therein). (1) They are exceedingly luminous, with
peak absolute blue magnitudes averaging $-19.2$ if the Hubble
constant, $H_0$, is 65 km s$^{-1}$ Mpc$^{-1}$. (2) ``Normal" SNe~Ia
have small dispersion among their peak absolute magnitudes ($\sigma
\lesssim 0.3$ mag). (3) Our understanding of the progenitors and
explosion mechanism of SNe~Ia is on a reasonably firm physical basis.
(4) Little cosmic evolution is expected in the peak luminosities of
SNe~Ia, and it can be modeled. This makes SNe~Ia superior to
galaxies as distance indicators. (5) One can perform {\it local} tests
of various possible complications and evolutionary effects by
comparing nearby SNe~Ia in different environments.

   Research on SNe~Ia in the 1990s has demonstrated their enormous
potential as cosmological distance indicators. Although there are
subtle effects that must indeed be taken into account, it appears that
SNe~Ia provide among the most accurate values of $H_0$, $q_0$ (the
deceleration parameter), $\Omega_M$ (the matter density), and
$\Omega_\Lambda$ (the cosmological constant, $\Lambda c^2/3H_0^2$).

   There are now two major teams involved in the systematic
investigation of high-redshift SNe~Ia for cosmological purposes. The
``Supernova Cosmology Project" (SCP) is led by Saul Perlmutter of the
Lawrence Berkeley Laboratory, while the ``High-Z Supernova Search
Team" (HZT) is led by Brian Schmidt of the Mt. Stromlo and Siding
Springs Observatories. One of us (A.V.F.) has worked with both teams,
but his primary allegiance is now with the HZT. In this paper
we present results from the HZT.

\section*{ HOMOGENEITY AND HETEROGENEITY}

   The traditional way in which SNe~Ia have been used for cosmological distance
determinations has been to assume that they are perfect ``standard candles" and
to compare their observed peak brightness with those of SNe~Ia in galaxies
whose distances have been independently determined (e.g., Cepheids). The
rationale is that SNe~Ia exhibit relatively little scatter in their peak blue
luminosity ($\sigma_B \approx 0.4$--0.5 mag; \cite{bra93}), and even less if
``peculiar" or highly reddened objects are eliminated from consideration by
using a color cut.  Moreover, the optical spectra of SNe~Ia are usually rather
homogeneous, if care is taken to compare objects at similar times relative to
maximum brightness (\cite{rie97} and references therein).  Over 80\% of all SNe~Ia discovered through the early 1990s
were ``normal" \cite{bfn93}.

   From a Hubble diagram constructed with unreddened, moderately
distant SNe~Ia ($z \lesssim 0.1$) for which peculiar motions should be
small and relative distances (as given by ratios of redshifts) are
accurate, Vaughan {\it et al.} \cite{vau95} find that

\begin{equation}
<M_B({\rm max})> \ = \ (-19.74 \pm 0.06) + 5\, {\rm log}\, (H_0/50)~{\rm mag}.
\end{equation}

\noindent In a series of papers, Sandage {\it et al.} \cite{san96} and 
Saha {\it et al.}
\cite{sah97} combine similar relations with {\it Hubble Space Telescope
(HST)} Cepheid distances to the host galaxies of seven SNe~Ia to derive
$H_0 = 57 \pm 4$ km s$^{-1}$ Mpc$^{-1}$.

   Over the past decade it has become clear, however, that SNe~Ia do {\it not}
constitute a perfectly homogeneous subclass (e.g., \cite{avf97b,avf97a}).  In
retrospect this should have been obvious: the Hubble diagram for SNe~Ia
exhibits scatter larger than the photometric errors, the dispersion actually
{\it rises} when reddening corrections are applied (under the assumption that
all SNe~Ia have uniform, very blue intrinsic colors at maximum;
\cite{van92,san93}), and there are some significant outliers whose anomalous
magnitudes cannot possibly be explained by extinction alone.

    Spectroscopic and photometric peculiarities have been noted with increasing
frequency in well-observed SNe~Ia. A striking case is SN 1991T; its pre-maximum
spectrum did not exhibit Si~II or Ca~II absorption lines, yet two months past
maximum brightness the spectrum was nearly indistinguishable from that of a
classical SN~Ia \cite{avf92b,phi92}.  The light curves of SN 1991T were
slightly broader than the SN~Ia template curves, and the object was probably
somewhat more luminous than average at maximum. The reigning champion of well
observed, peculiar SNe~Ia is SN 1991bg \cite{avf92a,lei93,tur96}.  At maximum
brightness it was subluminous by 1.6 mag in $V$ and 2.5 mag in $B$, its colors
were intrinsically red, and its spectrum was peculiar (with a deep absorption
trough due to Ti~II).  Moreover, the decline from maximum brightness was very
steep, the $I$-band light curve did not exhibit a secondary maximum like normal
SNe~Ia, and the velocity of the ejecta was unusually low. The photometric
heterogeneity among SNe~Ia is well demonstrated by Suntzeff \cite{sun96a} with
five objects having excellent $BVRI$ light curves.

\section*{ COSMOLOGICAL USES: LOW REDSHIFTS}

   Although SNe~Ia can no longer be considered perfect ``standard
candles," they are still exceptionally useful for cosmological
distance determinations. Excluding those of low luminosity (which are
hard to find, especially at large distances), most of the nearby SNe~Ia
that had been  discovered through the early 1990s were {\it
nearly} standard (\cite{bfn93}, but see Li {\it et al.} \cite{wli00a}
for recent evidence of a higher intrinsic peculiarity rate).
Also, after many tenuous suggestions (e.g., \cite{psk77,psk84,bra81}),
convincing evidence has finally been found for a {\it
correlation} between light-curve shape and luminosity. Phillips \cite{phi93}
achieved this by quantifying the photometric differences among a set
of nine well-observed SNe~Ia using a parameter, $\Delta m_{15}(B)$,
which measures the total drop (in $B$ magnitudes) from maximum to $t =
15$ days after $B$ maximum. In all cases the host galaxies of his
SNe~Ia have accurate relative distances from surface brightness
fluctuations or from the Tully-Fisher relation. In $B$, the SNe~Ia
exhibit a total spread of $\sim 2$ mag in maximum luminosity, and the
intrinsically bright SNe~Ia clearly decline more slowly than dim
ones. The range in absolute magnitude is smaller in $V$ and $I$,
making the correlation with $\Delta m_{15}(B)$ less steep than in $B$,
but it is present nonetheless.

   Using SNe~Ia discovered during the Cal\'an/Tololo survey ($z
\lesssim 0.1$), Hamuy {\it et al.} \cite{ham95,ham96b} confirm and refine the
Phillips \cite{phi93} correlation between $\Delta m_{15}(B)$ and
$M_{max}(B,V)$: it is not as steep as had been claimed. Apparently the
slope is steep only at low luminosities; thus, objects such as SN
1991bg skew the slope of the best-fitting single straight line. Hamuy
{\it et al.} reduce the scatter in the Hubble diagram of normal, unreddened
SNe~Ia to only 0.17 mag in $B$ and 0.14 mag in $V$; see also \cite{tri97}.

   In a similar effort, Riess, Press, \& Kirshner \cite{rpk95} show that the
luminosity of SNe~Ia correlates with the detailed shape of the overall light
curve. They form a ``training set" of light-curve shapes from 9 well-observed
SNe~Ia having known relative distances, including very peculiar objects (e.g.,
SN 1991bg). When the light curves of an independent sample of 13 SNe~Ia (the
Cal\'an/Tololo survey) are analyzed with this set of basis vectors, the
dispersion in the $V$-band Hubble diagram drops from 0.50 to 0.21 mag, and the
Hubble constant rises from $53 \pm 11$ to $67 \pm 7$ km s$^{-1}$ Mpc$^{-1}$,
comparable to the conclusions of Hamuy {\it et al.}  \cite{ham95,ham96b}. About half
of the rise in $H_0$ results from a change in the position of the ``ridge line"
defining the linear Hubble relation, and half is from a correction to the
luminosity of some of the local calibrators which appear to be unusually
luminous (e.g., SN 1972E).

   By using light-curve shapes measured through several different filters,
Riess, Press, \& Kirshner \cite{rpk96a} extend their analysis and objectively
eliminate the effects of interstellar extinction: a SN~Ia that has an unusually
red $B-V$ color at maximum brightness is assumed to be {\it intrinsically}
subluminous if its light curves rise and decline quickly, or of normal
luminosity but significantly {\it reddened} if its light curves rise and
decline slowly. With a set of 20 SNe~Ia consisting of the Cal\'an/Tololo sample
and their own objects, they show that the dispersion decreases from 0.52 mag to
0.12 mag after application of this ``multi-color light curve shape" (MLCS)
method. The results from a recent, expanded set of nearly 50 SNe~Ia
indicate that the dispersion decreases from 0.44 mag to 0.15 mag (Figure
1). The resulting Hubble constant is $65 \pm 2$ (statistical) km s$^{-1}$
Mpc$^{-1}$, with an additional systematic and zero-point uncertainty of $\pm 5$
km s$^{-1}$ Mpc$^{-1}$. Riess {\it et al.}  \cite{rpk96a} also show that the Hubble
flow is remarkably linear; indeed, SNe~Ia now constitute the best evidence for
linearity. Finally, they argue that the dust affecting SNe~Ia is {\it not} of
circumstellar origin, and show quantitatively that the extinction curve in
external galaxies typically does not differ from that in the Milky Way
(cf. \cite{bra92}, but see \cite{tri98}).

   The advantage of systematically correcting the luminosities of
SNe~Ia at high redshifts rather than trying to isolate ``normal"
ones seems clear in view of evidence that the luminosity of
SNe~Ia may be a function of stellar population. If the most luminous
SNe~Ia occur in young stellar populations \cite{ham95,ham96a,bra96},
then we might expect the mean peak
luminosity of high-redshift SNe~Ia to differ from that of a local
sample. Alternatively, the use of Cepheids (Population I objects) to
calibrate local SNe~Ia can lead to a zero point that is too luminous.
On the other hand, as long as the physics of SNe~Ia is essentially the
same in young stellar populations locally and at high redshift, we
should be able to adopt the luminosity correction methods (photometric 
and spectroscopic) found from detailed studies of low-redshift
SNe~Ia.

\smallskip

\hbox{
\hskip +1.0truein
\vbox{\hsize 3.0 truein
\psfig{figure=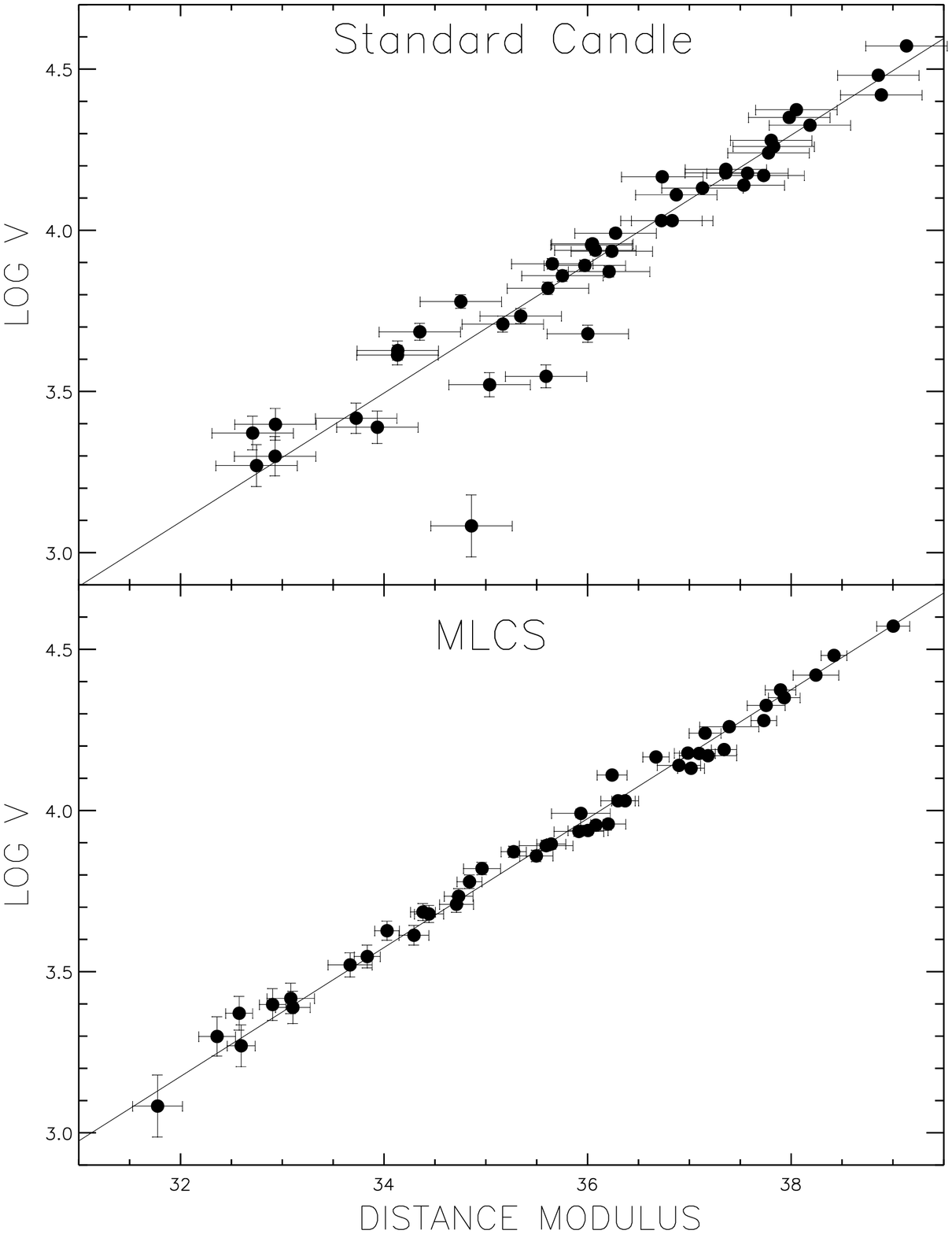,height=4.0truein,angle=0}
}
}

\bigskip
\bigskip

\noindent
{\it Figure 1:} Hubble diagrams for SNe~Ia \cite{rie00b} with velocities
(km s$^{-1}$) in the COBE rest frame on the Cepheid distance scale. {\it Top:} The
objects are assumed to be {\it standard candles} and there is no correction for
extinction; the result is $\sigma = 0.42$ mag and $H_0 = 58 \pm 8$ km s$^{-1}$
Mpc$^{-1}$. {\it Bottom:} The same objects, after correction for
reddening and intrinsic differences in luminosity. The result is $\sigma =
0.15$ mag and $H_0 = 65 \pm 2$ (statistical) km s$^{-1}$ Mpc$^{-1}$.

\bigskip

   Large numbers of nearby SNe~Ia are now being found by the Lick Observatory
Supernova Search (LOSS) conducted with the 0.76-m Katzman Automatic Imaging
Telescope (KAIT; \cite{wli00a,avf00a}). CCD images are taken of about 1000
galaxies per night and compared with KAIT ``template images" obtained earlier;
the templates are automatically subtracted from the new images and analyzed
with computer software.  The system reobserves the best candidates the same
night, to eliminate star-like cosmic rays, asteroids, and other sources of
false alarms. The next day, undergraduate students at UC Berkeley examine all
candidates, including weak ones, and they glance at all subtracted images to
locate SNe that might be near bright, poorly subtracted stars or galactic
nuclei. LOSS discovered 20 SNe (of all types) in 1998 and 40 SNe in 1999,
making it the world's most successful search for nearby SNe. The most important
objects were photometrically monitored through $BVRI$ (and sometimes $U$)
filters (e.g., \cite{wli00b}), and unfiltered follow-up observations were made
of all of them during the course of the SN search. This growing sample of
well-observed SNe~Ia should allow us to more precisely calibrate the MLCS
method, as well as to look for correlations between the observed properties of
the SNe and their environment (Hubble type of host galaxy, metallicity, stellar
population, etc.).

\section*{ COSMOLOGICAL USES: HIGH REDSHIFTS}

   These same techniques can be applied to construct a Hubble diagram
with high-redshift SNe~Ia, from which the value of $q_0$ can be
determined. With enough objects spanning a range of redshifts, we can
determine $\Omega_M$ and $\Omega_\Lambda$ independently (e.g., 
\cite{goo95}). Contours of peak apparent $R$-band magnitude 
for SNe~Ia at two redshifts have different slopes in the
$\Omega_M$--$\Omega_\Lambda$ plane, and the regions of intersection
provide the answers we seek.

\subsection*{ The Search}

   Based on the pioneering work of Norgaard-Nielsen {\it et al.} \cite{nor89},
whose goal was to find SNe in moderate-redshift clusters of galaxies,
the SCP \cite{per97} and our HZT \cite{sch98} devised a
strategy that almost guarantees the discovery of many faint, distant
SNe~Ia on demand, during a predetermined set of nights.  This ``batch"
approach to studying distant SNe allows follow-up spectroscopy and
photometry to be {\it scheduled} in advance, resulting in a systematic
study not possible with random discoveries.  Most of the searched
fields are equatorial, permitting follow-up from both hemispheres.

    Our approach is simple in principle; see \cite{sch98} for
details, and for a description of our first high-redshift SN~Ia (SN
1995K). Pairs of first-epoch images are obtained with the CTIO or CFHT
4-m telescopes and wide-angle imaging cameras during the nights just
after new moon, followed by second-epoch images 3--4 weeks later.
(Pairs of images permit removal of cosmic rays, asteroids, and distant
Kuiper-belt objects.) These are compared immediately using well-tested
software, and new SN candidates are identified in the second-epoch
images (Figure 2). Spectra are obtained as soon as possible after discovery to
verify that the objects are SNe~Ia and determine their redshifts.
Each team has already found over 80 SNe in concentrated batches, as
reported in numerous {\it IAU Circulars} (e.g., \cite{per95},
11 SNe with $0.16 \lesssim z \lesssim 0.65$; \cite{sun96b},
17 SNe with $0.09 \lesssim z \lesssim 0.84$).

   Intensive photometry of the SNe~Ia commences within a few days after
procurement of the second-epoch images; it is continued throughout the ensuing
and subsequent dark runs. In a few cases {\it HST} images are obtained. As
expected, most of the discoveries are {\it on the rise or near maximum
brightness}.  When possible, the SNe are observed in filters which closely
match the redshifted $B$ and $V$ bands; this way, the $K$-corrections become
only a second-order effect \cite{kim96}. Custom-designed filters for redshifts
centered on 0.35 and 0.45 are used by our HZT \cite{sch98}, when appropriate.
We try to obtain excellent {\it multi-color} light curves, so that reddening
and luminosity corrections can be applied \cite{rpk96a,ham96a,ham96b}.

\medskip

\hbox{
\hskip +0.7truein
\vbox{\hsize 4.0 truein
\psfig{figure=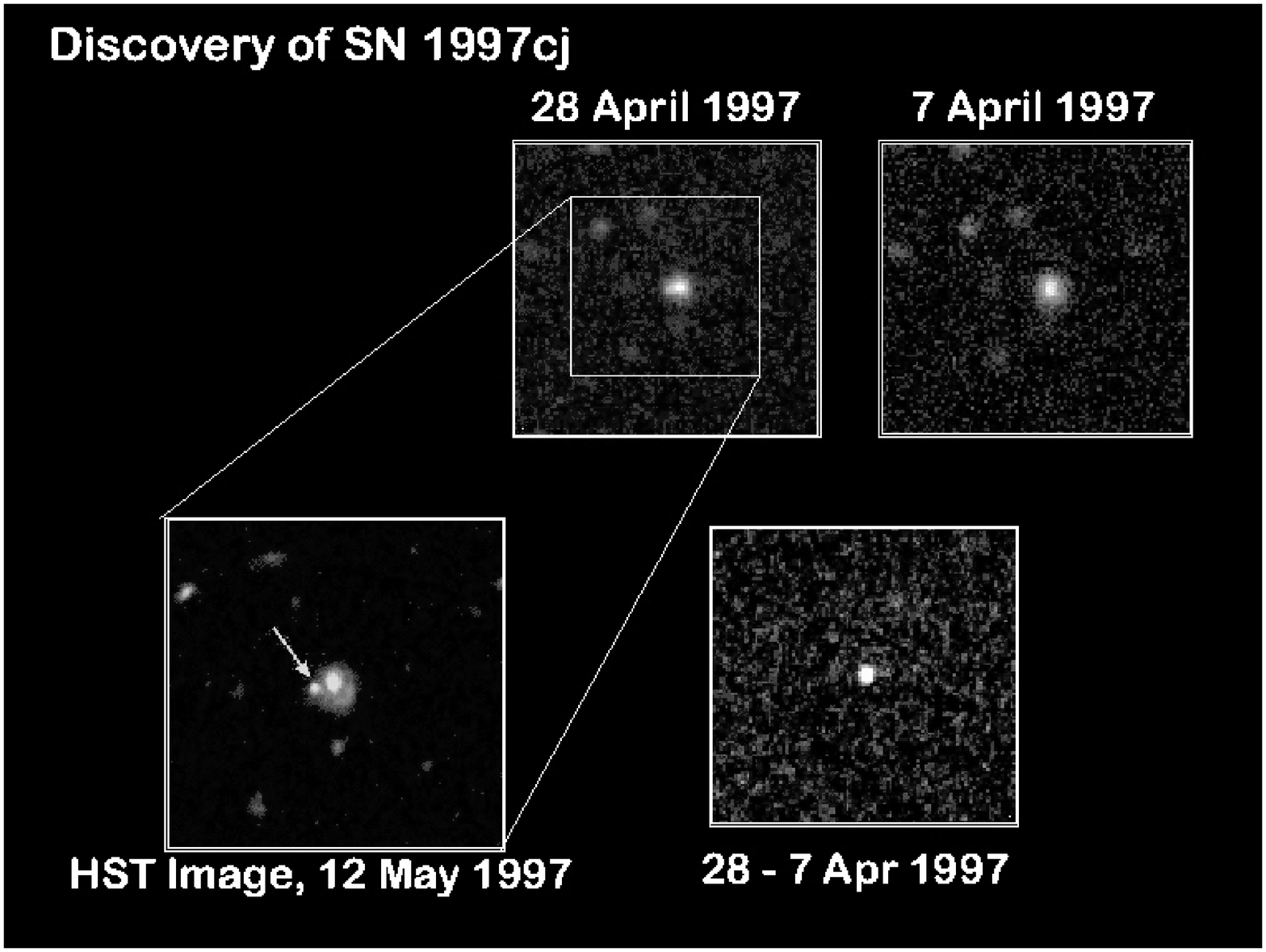,height=4.0truein,angle=270}
}
}

\bigskip

\noindent
{\it Figure 2:} Discovery image of SN 1997cj (28 April 1997), along 
with the template image (7 April 1997) and an {\it HST} image 
obtained subsequently. The net (subtracted) image is also shown.

\bigskip

  Although SNe in the magnitude range 22--22.5 can sometimes be
spectroscopically confirmed with 4-m class telescopes, the signal-to-noise
ratios are low, even after several hours of integration. Certainly Keck is
required for the fainter objects (mag 22.5--24.5). With Keck, not only can we
rapidly confirm a large number of candidate SNe, but we can search for
peculiarities in the spectra that might indicate evolution of SNe~Ia with
redshift.  Moreover, high-quality spectra allow us to measure the age of a
supernova: we have developed a method for automatically comparing the spectrum
of a SN~Ia with a library of spectra corresponding to many different epochs in
the development of SNe~Ia \cite{rie97}.  Our technique also has great practical
utility at the telescope: we can determine the age of a SN ``on the fly,''
within half an hour after obtaining its spectrum. This allows us to rapidly
decide which SNe are best for subsequent photometric follow-up, and we
immediately alert our collaborators on other telescopes.

\subsection*{ Results}

   First, we note that the light curves of high-redshift SNe~Ia are
broader than those of nearby SNe~Ia; the initial indications
\cite{lei96,gol97} are amply
confirmed with our larger samples. Quantitatively, the amount by which
the light curves are ``stretched'' is consistent with a factor of $1 +
z$, as expected if redshifts are produced by the expansion of space
rather than by ``tired light.'' We were also able to demonstrate this
{\it spectroscopically} at the $2\sigma$ confidence level for a single
object: the spectrum of SN 1996bj ($z = 0.57$) evolved more slowly
than those of nearby SNe~Ia, by a factor consistent with $1 + z$
\cite{rie97}. More recently, we have used observations of SN
1997ex ($z = 0.36$) at three epochs to conclusively verify the effects
of time dilation: temporal changes in the spectra are slower than
those of nearby SNe~Ia by roughly the expected factor of 1.36 \cite{avf00b}.

   Following our Spring 1997 campaign, in which we found a SN with $z
= 0.97$ (SN 1997ck), and for which we obtained {\it HST}
follow-up images of three SNe, we published our first substantial results
concerning the density of the Universe \cite{gar98a}:
$\Omega_M = 0.35 \pm 0.3$ under the {\it assumption} that $\Omega_{\rm
total} = 1$, or $\Omega_M = -0.1 \pm 0.5$ under the {\it assumption}
that $\Omega_\Lambda = 0$.  Our independent analysis of 10 SNe~Ia
using the ``snapshot'' distance method (with which conclusions are
drawn from sparsely observed SNe~Ia) gives quantitatively similar
conclusions \cite{rie98a}.

\bigskip

\hbox{
\hskip +0truein
\vbox{\hsize 2.5 truein
\psfig{figure=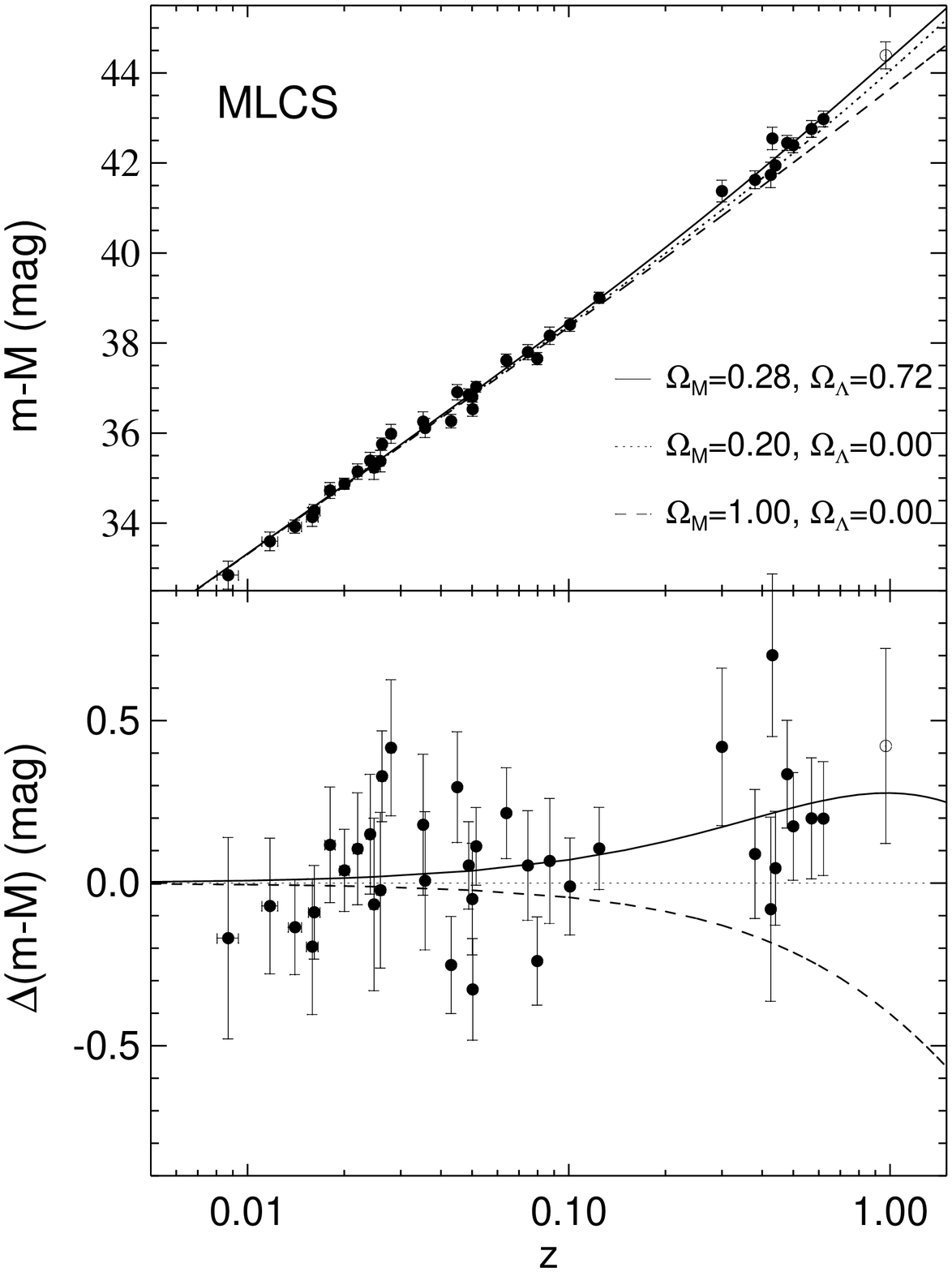,height=3truein,angle=0}
}
\hskip +.1truein
\vbox{\hsize 3.3 truein
\psfig{figure=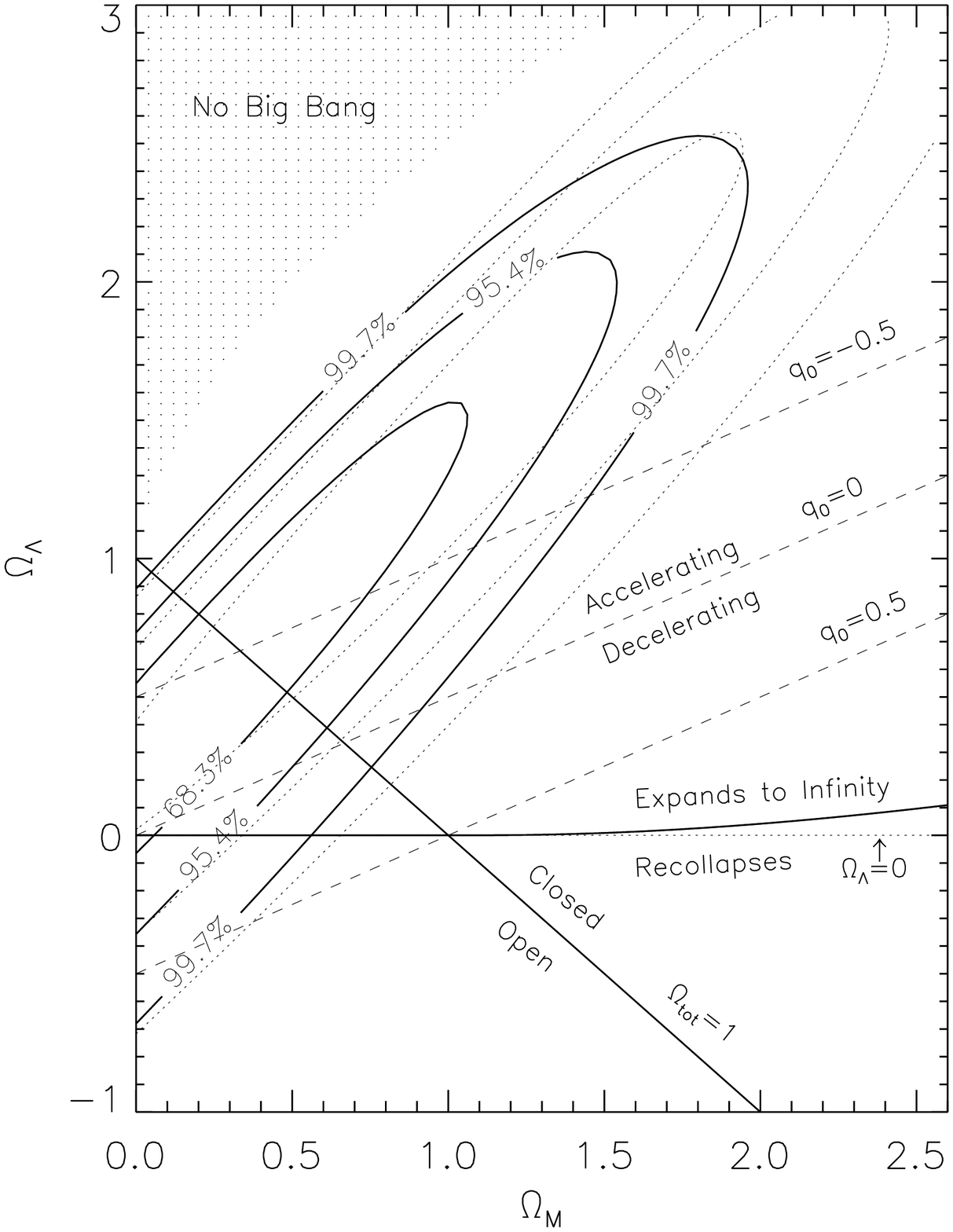,height=3.0truein,angle=0}
}
}

\bigskip
\medskip

\noindent
{\it Figure 3 (left):} The upper panel shows the Hubble diagram for the
low-redshift and high-redshift SN~Ia samples with distances measured
from the MLCS method; see \cite{rie98b}. Overplotted are 
three world models: ``low'' and ``high'' $\Omega_M$ with 
$\Omega_\Lambda=0$, and the best fit for a flat universe 
($\Omega_M=0.28$, $\Omega_\Lambda=0.72$).  The bottom
panel shows the difference between data and models from the
$\Omega_M=0.20$, $\Omega_\Lambda=0$ prediction.
Except for SN 1997ck ({\it open symbol}; $z = 0.97$), which lacks 
spectroscopic confirmation and was excluded from the fit, 
only the 9 best-observed high-redshift SNe~Ia are shown.
The average difference between the data and the
$\Omega_M=0.20$, $\Omega_\Lambda=0$ prediction is 0.25 mag.

\noindent
{\it Figure 4 (right):} Joint confidence intervals for
($\Omega_M$,$\Omega_\Lambda$) from SNe~Ia \cite{rie98b}.
The solid contours are
results from the MLCS method applied to 10 well-observed SN~Ia light
curves, together with the snapshot method \cite{rie98a} applied
to 6 incomplete SN~Ia light curves.  The dotted contours are for the
same objects excluding SN 1997ck ($z=0.97$).  Regions representing
specific cosmological scenarios are illustrated.

\bigskip

   Our next results, obtained from a total of 16 high-$z$ SNe~Ia, were
announced at a conference in February 1998 \cite{avf98} and formally published
by Riess {\it et al.} \cite{rie98b} in September 1998. The Hubble diagram (from a
refined version of the MLCS method \cite{rie98b}) for the 10 best-observed
high-$z$ SNe~Ia is given in Figure 3, while Figure 4 illustrates the derived
confidence contours in the $\Omega_M$--$\Omega_\Lambda$ plane. We confirm our
previous suggestion that $\Omega_M$ is low. Even more exciting, however, is our
conclusion that $\Omega_\Lambda$ is {\it nonzero} at the 3$\sigma$ statistical
confidence level. With the MLCS method applied to the full set of 16 SNe~Ia,
our formal results are $\Omega_M = 0.24 \pm 0.10$ if $\Omega_{\rm total} = 1$,
or $\Omega_M = -0.35 \pm 0.18$ (unphysical) if $\Omega_\Lambda = 0$. If we
demand that $\Omega_M = 0.2$, then the best value for $\Omega_\Lambda$ is $0.66
\pm 0.21$.  These conclusions do not change significantly if only the 9
best-observed SNe~Ia are used (Figure 3; $\Omega_M = 0.28 \pm 0.10$ if
$\Omega_{\rm total} = 1$).  The $\Delta m_{15}(B)$ method yields similar
results; if anything, the case for a positive cosmological constant
strengthens. (For brevity, in this paper we won't quote the $\Delta m_{15}(B)$
numbers; see \cite{rie98b} for details.)  From an essentially independent set
of 42 high-$z$ SNe~Ia (only 2 objects in common), the SCP obtains almost
identical results \cite{per99}. This suggests that neither team has made a
large, simple blunder!

   Recently, we have calibrated an additional sample of 9 high-$z$ SNe~Ia,
including several observed with {\it HST}.  Preliminary analysis suggests that
the new data are entirely consistent with the old results, thereby
strengthening their statistical significance. Figure 5 shows the tentative
Hubble diagram; full details will be published elsewhere.

   Though not drawn in Figure 4, the expected confidence contours from
measurements of the angular scale of the first acoustic peak of the cosmic
microwave background radiation (CMBR) are nearly perpendicular to those
provided by SNe~Ia (e.g., \cite{zal97,eis98}); thus, the two techniques provide
complementary information. A stunning result was already available by mid-1998
from existing measurements \cite{han98,lin98}: our analysis of the data in
\cite{rie98b} demonstrates that $\Omega_M + \Omega_\Lambda = 0.94 \pm 0.26$,
when the SN and CMBR constraints are combined \cite{gar98b} (see also
\cite{lin98,efs99}, and others). As shown in Figure 6, the confidence contours
are nearly circular, instead of highly eccentric ellipses as in Figure 4.  Just
a few days before the Second Tropical Workshop, the more accurate and precise
results of the BOOMERANG collaboration were announced \cite{deb00}, and shortly
thereafter the MAXIMA collaboration distributed their very similar findings
\cite{han00,bal00}; the TOCO measurements \cite{mil99} are also relevant. The
bottom line is that we appear to live in a flat universe: $\Omega_{\rm total} =
1$. Combined with the supernova results, the evidence for nonzero
$\Omega_\Lambda$ is strong. Making the argument even more compelling is the
fact that various studies of clusters of galaxies (see summary by \cite{bah99})
show that $\Omega_M \approx 0.3$, so if the CMBR results are correct, one is
led to the independent conclusion that $\Omega_\Lambda > 0$.  We eagerly look
forward to future CMBR measurements of even greater precision.

   The dynamical age of the Universe can be calculated from the
cosmological parameters. In an empty Universe with no cosmological
constant, the dynamical age is simply the ``Hubble time" (i.e., the
inverse of the Hubble constant); there is no deceleration.  SNe~Ia
yield $H_0 = 65 \pm 2$ km s$^{-1}$ Mpc$^{-1}$ (statistical uncertainty
only), and a Hubble time of $15.1 \pm 0.5$ Gyr. For a more complex
cosmology, integrating the velocity of the expansion from the current
epoch ($z=0$) to the beginning ($z=\infty$) yields an expression for
the dynamical age. As shown in detail by Riess {\it et al.} \cite{rie98b}, we
obtain a value of 14.2$^{+1.0}_{-0.8}$ Gyr using the likely range for
$(\Omega_M, \Omega_\Lambda)$ that we measure.  (The precision is so
high because our experiment is sensitive to roughly the {\it difference}
between $\Omega_M$ and $\Omega_\Lambda$, and the dynamical age also
varies in approximately this way.)  Including the {\it systematic}
uncertainty of the Cepheid distance scale, which may be up to 10\%, a
reasonable estimate of the dynamical age is $14.2 \pm 1.7$ Gyr.

\bigskip

\hbox{
\hskip +.0truein
\vbox{\hsize 2.5 truein
\psfig{figure=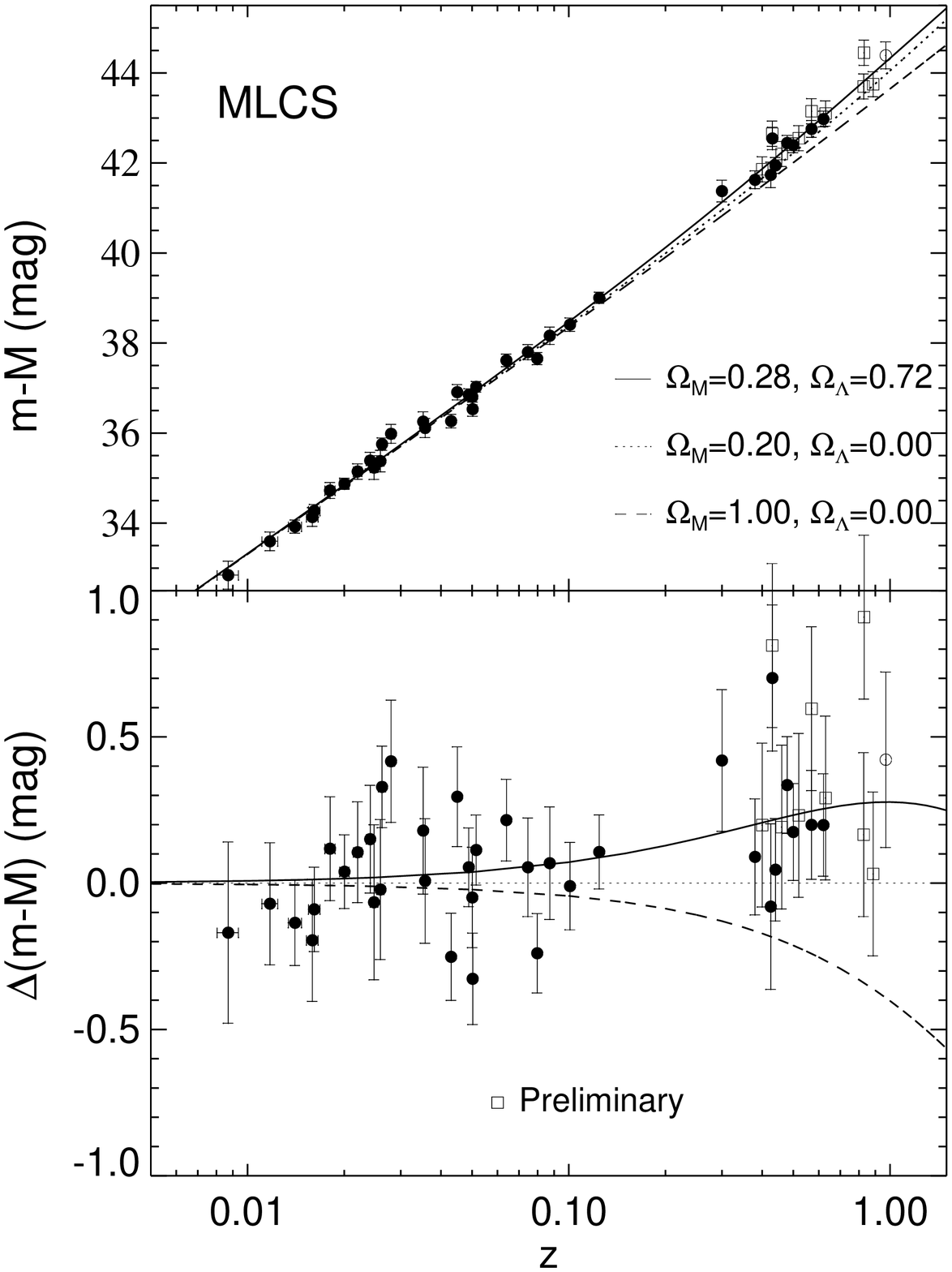,height=3truein,angle=0}
}
\hskip +.4truein
\vbox{\hsize 3.2 truein
\psfig{figure=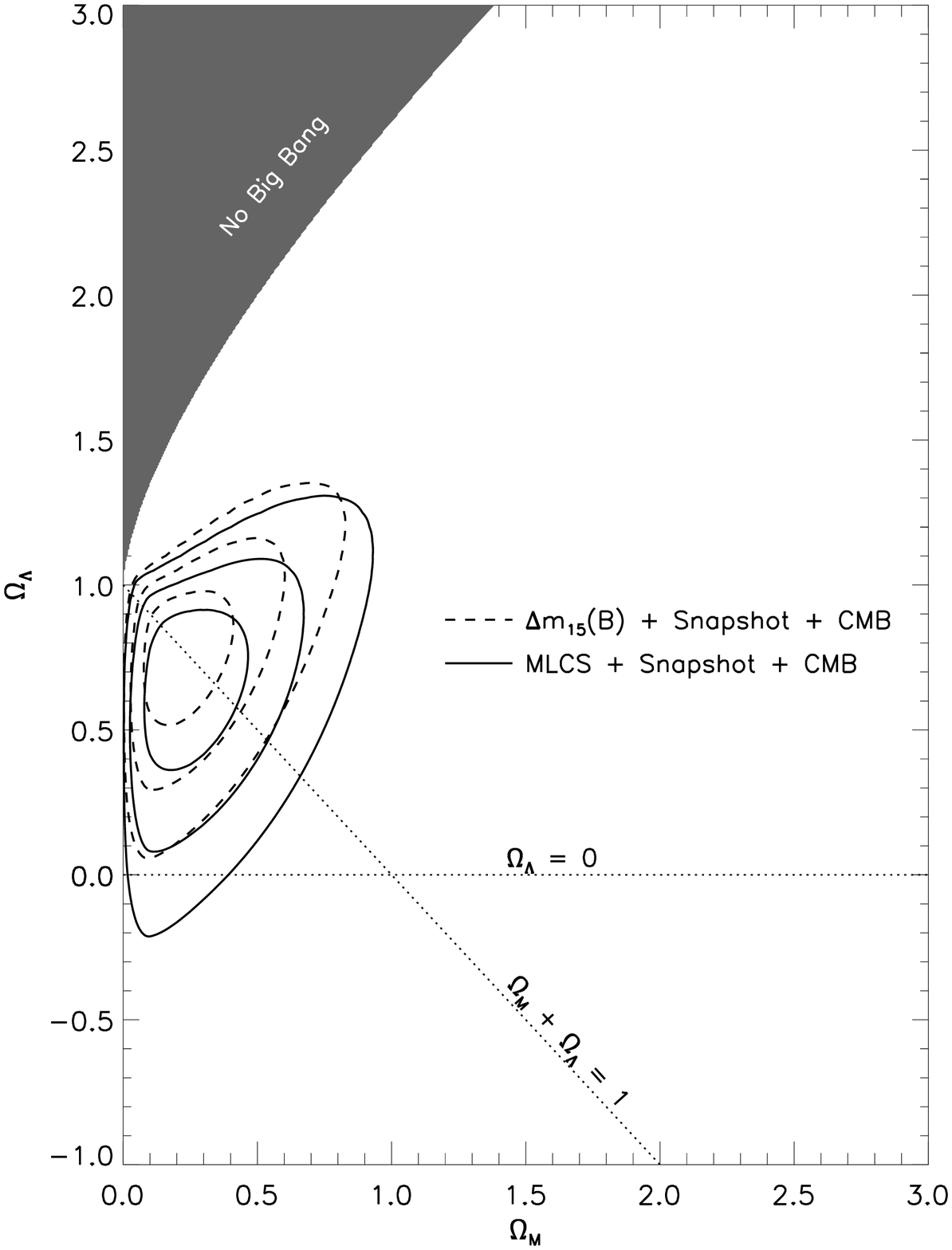,height=3truein,angle=0}
}
}
\bigskip

\noindent
{\it Figure 5 (left):}
As in Figure 3, the upper panel shows the Hubble diagram for the
low-$z$ and high-$z$ SN~Ia samples. Here, we include preliminary analysis of 
9 additional SNe~Ia {\it (open squares)}. The bottom
panel shows the difference between data and models from the
$\Omega_M=0.20$, $\Omega_\Lambda=0$ prediction.

\noindent
{\it Figure 6 (right):} 
The HZT's combined constraints from SNe~Ia (Figure 3) and the position of
the first acoustic peak of the CMB angular power spectrum, based on data
available in mid-1998; see \cite{gar98b}.
The contours mark the 68\%, 95.4\%, and 99.7\% enclosed probability
regions. Solid curves correspond to results from the MLCS method, while 
dotted ones are from 
the $\Delta m_{15}(B)$ method; all 16 SNe~Ia 
in \cite{rie98b} were used.

\bigskip

  This result is consistent with ages determined from various other techniques
such as the cooling of white dwarfs (Galactic disk $> 9.5$ Gyr \cite{osw96}),
radioactive dating of stars via the thorium and europium abundances ($15.2 \pm
3.7$ Gyr \cite{cow97}), and studies of globular clusters (10--15 Gyr, depending
on whether {\it Hipparcos} parallaxes of Cepheids are adopted
\cite{gra97,cha98}).  Evidently, there is no longer a problem that the age of
the oldest stars seems greater than the dynamical age of the Universe.

\section*{ DISCUSSION}

   {\it High-redshift SNe~Ia are observed to be dimmer than expected
in an empty Universe (i.e., $\Omega_M=0$) with no cosmological
constant.}  A cosmological explanation for this observation is that a
positive vacuum energy density accelerates the expansion.  Mass
density in the Universe exacerbates this problem, requiring even more
vacuum energy.  For a Universe with $\Omega_M=0.2$, the average MLCS
distance moduli of the well-observed SNe are 0.25 mag larger (i.e.,
12.5\% greater distances) than the prediction from $\Omega_\Lambda=0$.
The average MLCS distance moduli are still 0.18 mag bigger than
required for a 68.3\% (1$\sigma$) consistency in a universe with
$\Omega_M=0.2$ and without a cosmological constant.  The derived value
of $q_0$ is $-0.75 \pm 0.32$, implying that the expansion of the
Universe is accelerating. If $\Omega_\Lambda$ really is constant, then at
least the region of the Universe we have observed ($z \lesssim 0.8$) will 
expand eternally. Under the simplifying assumption of global homogeneity
and isotropy, the entire Universe will behave in this manner.

   A very important point is that the dispersion in the peak
luminosities of SNe~Ia ($\sigma = 0.15$ mag) is low after application
of the MLCS method of \cite{rpk96a,rie98b}. With 16 SNe~Ia, our
effective uncertainty is $0.15/4 \approx 0.04$ mag, less than the
expected difference of 0.25 mag between universes with $\Omega_\Lambda
= 0$ and 0.76 (and low $\Omega_M$); see Figure 3.  Systematic
uncertainties of even 0.05 mag (e.g., in the extinction) are
significant, and at 0.1 mag they dominate any decrease in statistical
uncertainty gained with a larger sample of SNe~Ia.  Thus, our
conclusions with only 16 SNe~Ia are already limited by systematic
uncertainties, {\it not} by statistical uncertainties ---  but of
course the 9 new objects further strengthen our case.
  
   Here we explore the major possible systematic effects that could invalidate
our results. Of those that can be quantified at the present time, none appears
to reconcile the data with $\Omega_\Lambda = 0$, though further work is
necessary to verify this. Additional details can be found in \cite{sch98} and
especially \cite{rie98b}.

\subsection*{ Evolution}

   Perhaps the most obvious possible culprit is {\it evolution} of SNe~Ia over
cosmic time, due to changes in metallicity, progenitor mass, or some other
factor. If the peak luminosity of SNe~Ia were lower at high redshift, then
the case for $\Omega_\Lambda > 0$ would weaken.  Conversely, if the distant
explosions are more powerful, then the case for acceleration
strengthens. Theorists are not yet sure what the sign of the effect will be, if
it's present at all; different assumptions lead to different conclusions
\cite{hof98,ume99,dom99,yun00,nom00}.

     Of course, it is very difficult to obtain an independent measure of the
peak luminosity of high-$z$ SNe~Ia, and hence to directly test for luminosity
evolution. However, we can more easily determine whether {\it other} observable
properties of low-$z$ and high-$z$ SNe~Ia differ. If they are all the same, it
is more probable that the peak luminosity is constant as well --- but if they
differ, then the peak luminosity might also be affected (e.g., \cite{hof98}).
Drell {\it et al.} \cite{dre00}, for example, argue that there are reasons to
suspect evolution, because the average properties of existing samples of
high-$z$ and low-$z$ SNe~Ia seem to differ (e.g., the high-$z$ SNe~Ia are more
uniform).

   The local sample of SNe~Ia displays a weak correlation between
light-curve shape (or luminosity) and host galaxy type, in the sense
that the most luminous SNe~Ia with the broadest light curves only
occur in late-type galaxies.  Both early-type and late-type galaxies
provide hosts for dimmer SNe~Ia with narrower light curves \cite{ham96a}.  
The mean luminosity difference for SNe~Ia in late-type
and early-type galaxies is $\sim 0.3$ mag.  In addition, the SN~Ia
rate per unit luminosity is almost twice as high in late-type galaxies
as in early-type galaxies at the present epoch \cite{cap97}.
These results may indicate an evolution of SNe~Ia with
progenitor age. Possibly relevant physical parameters are the mass,
metallicity, and C/O ratio of the progenitor \cite{hof98}.

   We expect that the relation between light-curve shape and luminosity
that applies to the range of stellar populations and progenitor ages
encountered in the late-type and early-type hosts in our nearby sample
should also be applicable to the range we encounter in our distant
sample.  In fact, the range of age for SN~Ia progenitors in the nearby
sample is likely to be {\it larger} than the change in mean progenitor
age over the 4--6 Gyr lookback time to the high-$z$ sample.  Thus, to
first order at least, our local sample should correct our distances
for progenitor or age effects.

   We can place empirical constraints on the effect that a change in the
progenitor age would have on our SN~Ia distances by comparing
subsamples of low-redshift SNe~Ia believed to arise from old and young
progenitors.  In the nearby sample, the mean difference between the
distances for the early-type (8 SNe~Ia) and late-type hosts (19
SNe~Ia), at a given redshift, is 0.04 $\pm$ 0.07 mag from the MLCS
method.  This difference is consistent with zero.  Even if the SN~Ia
progenitors evolved from one population at low redshift to the other
at high redshift, we still would not explain the surplus in mean
distance of 0.25 mag over the $\Omega_\Lambda=0$ prediction.  

  Moreover, it is reassuring that initial comparisons of high-redshift SN~Ia
spectra appear remarkably similar to those observed at low redshift.  For
example, the spectral characteristics of SN 1998ai ($z = 0.49$) appear to be
essentially indistinguishable from those of normal low-redshift SNe~Ia; see
Figure 7. In fact, the most obviously discrepant spectrum in this figure is the
second one from the top, that of SN 1994B ($z = 0.09$); it is intentionally
included as a ``decoy" that illustrates the degree to which even the spectra of
nearby, relatively normal SNe~Ia can vary. Nevertheless, it is important to
note that a dispersion in luminosity (perhaps 0.2 mag) exists even among the
other, more normal SNe~Ia shown in Figure 7; thus, our spectra of SN 1998ai and
other high-redshift SNe~Ia are not yet sufficiently good for independent, {\it
precise} determinations of luminosity from spectral features \cite{nug95}.
Many of them, however, are sufficient for ruling out other supernovae types
(Figure 8), or for identifying gross peculiarities such as those shown by SNe
1991T and 1991bg; see Coil {\it et al.} \cite{coi00}.

  We can help verify that the SNe at $z \approx 0.5$ being used for cosmology
do not belong to a subluminous population of SNe~Ia by examining restframe
$I$-band light curves. Normal, nearby SNe~Ia show a pronounced second maximum
in the $I$ band about a month after the first maximum and typically about 0.5
mag fainter (e.g., \cite{for93,sun96a}). Subluminous SNe~Ia, in contrast, do not
show this second maximum, but rather follow a linear decline or show a muted
second maximum \cite{avf92a}. As discussed by Riess {\it et al.} \cite{rie00a}, some
evidence for the second maximum is seen from our existing $J$-band (restframe
$I$-band) data on SN 1999Q ($z = 0.46$); see Figure 9.  However, better data on
more SNe~Ia are needed to confirm the effect.

\bigskip

\hbox{
\hskip -0.3truein
\vbox{\hsize 2.5 truein
\psfig{figure=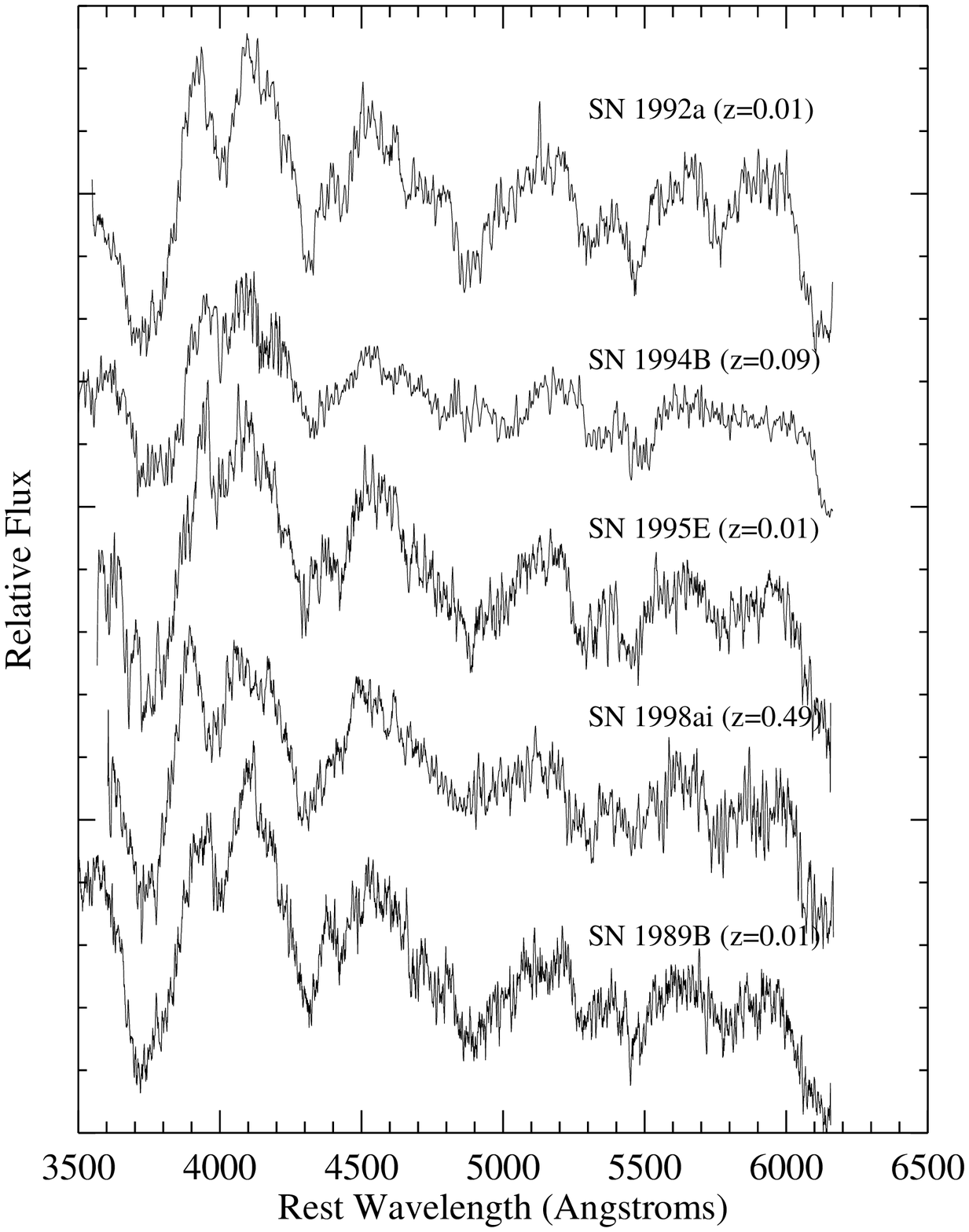,height=3.5truein,angle=0}
}
\hskip 0.4truein
\vbox{\hsize 2.5 truein
\psfig{figure=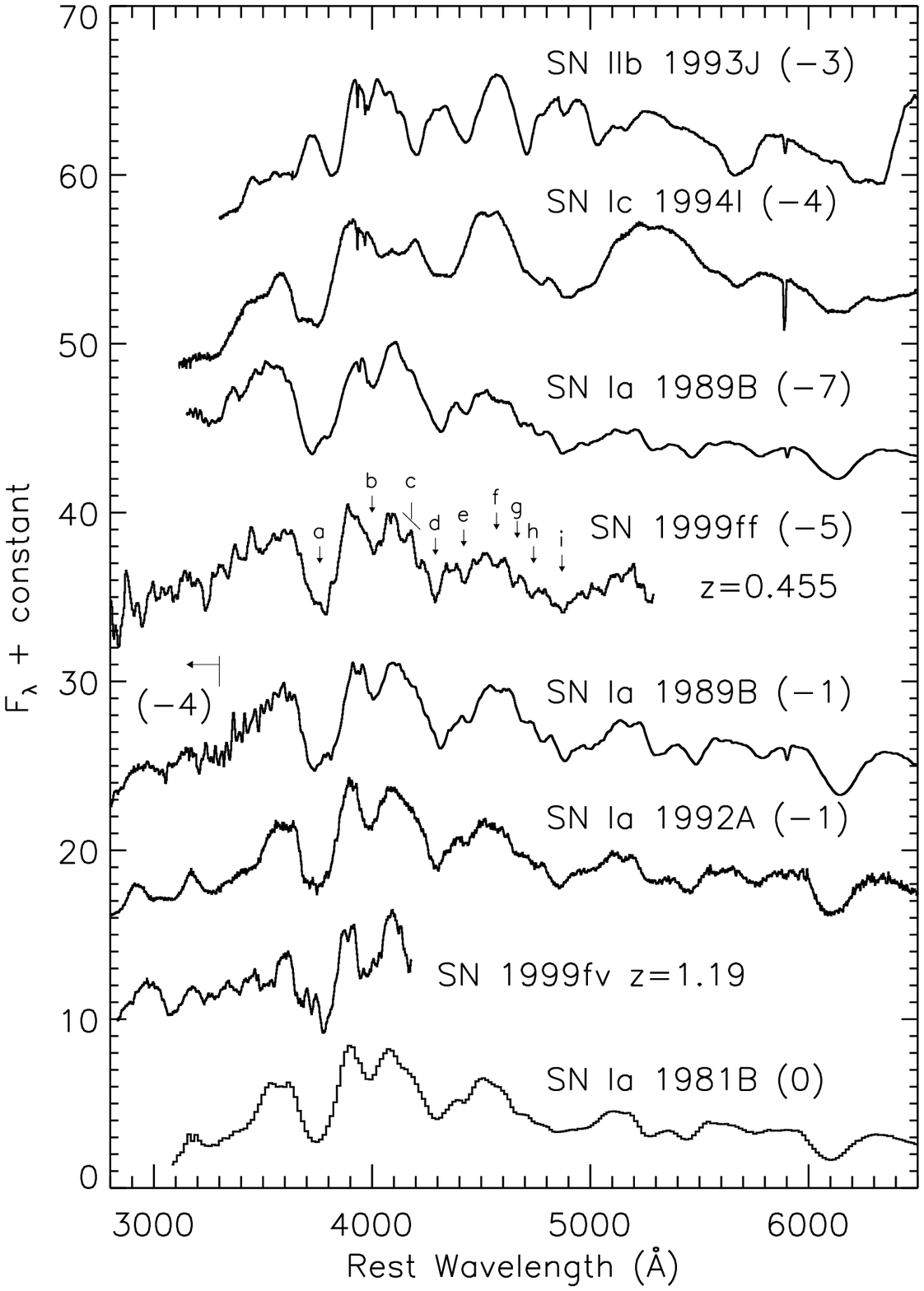,height=3.5truein,angle=0}
}
}

\bigskip

\noindent
{\it Figure 7 (left):}
Spectral comparison (in $f_{\lambda}$) of SN 1998ai ($z = 0.49$;
Keck spectrum) with low-redshift ($z < 0.1$) SNe~Ia at a similar age
($\sim 5$ days before maximum brightness), from \cite{rie98b}.
The spectra of the low-redshift 
SNe~Ia were resampled and convolved with Gaussian noise to match the 
quality of the spectrum of SN 1998ai. Overall, the agreement
in the spectra is excellent, tentatively suggesting that distant SNe~Ia are
physically similar to nearby SNe~Ia.  SN 1994B ($z = 0.09$) differs
the most from the others, and was included as a ``decoy."

\noindent
{\it Figure 8 (right):} Heavily smoothed spectra of two high-$z$ SNe (SN 1999ff at $z =
0.455$ and SN 1999fv at $z = 1.19$; quite noisy below $\sim$3500~\AA) are
presented along with several low-$z$ SN Ia spectra (SNe 1989B, 1992A, and
1981B), a SN Ib spectrum (SN 1993J), and a SN~Ic spectrum (SN 1994I); see
\cite{avf97b} for a discussion of spectra of various types of SNe. The date
of the spectra relative to $B$-band maximum is shown in parentheses after each
object's name. Specific features seen in SN 1999ff and labeled with a letter
are discussed by Coil {\it et al.} \cite{coi00}. This comparison shows that the two
high-$z$ SNe are most likely SNe~Ia.

\bigskip

  Another way of using light curves to test for possible evolution of SNe~Ia is
to see whether the rise time (from explosion to maximum brightness) is the same
for high-$z$ and low-$z$ SNe~Ia; a difference might indicate that the peak
luminosities are also different \cite{hof98}. We recently measured the risetime
of nearby SNe~Ia, using data from KAIT, the Beijing Astronomical Observatory
(BAO) SN search, and a few amateur astronomers \cite{rie99b}.  Though the exact
value of the risetime is a function of peak luminosity, for typical low-$z$
SNe~Ia we find $20.0 \pm 0.2$ days. We pointed out \cite{rie99a} that this
differs by $5.8\sigma$ from the {\it preliminary} risetime of $17.5 \pm 0.4$
days reported in conferences by the SCP \cite{gol98a,gol98b,gro98}. However, a
more thorough analysis of the SCP data \cite{ald00} shows that the high-$z$
uncertainty of $\pm 0.4$ days that the SCP originally reported was much too
small because it did not account for systematic effects. The revised
discrepancy with the low-$z$ risetime is about $2\sigma$ or less. Thus,
the apparent difference in risetimes might be insignificant. Even
if the difference is real, however, its relevance to the peak luminosity is
unclear; the light curves may differ only in the first few days after the
explosion, and this could be caused by small variations in conditions near the
outer part of the exploding white dwarf that are inconsequential at the peak.

\subsection*{ Extinction}
 
   Our SN~Ia distances have the important advantage of including
corrections for interstellar extinction occurring in the host galaxy
and the Milky Way. Extinction corrections based on the relation
between SN~Ia colors and luminosity improve distance precision for a
sample of nearby SNe~Ia that includes objects with substantial
extinction \cite{rpk96a}; the scatter in the
Hubble diagram is much reduced.  Moreover, the consistency of the
measured Hubble flow from SNe~Ia with late-type and early-type hosts
(see above) shows that the extinction corrections applied to dusty
SNe~Ia at low redshift do not alter the expansion rate from its value
measured from SNe~Ia in low dust environments. 

   In practice, our high-redshift SNe~Ia appear to suffer negligible
extinction; their $B-V$ colors at maximum brightness are normal, suggesting
little color excess due to reddening. Riess, Press, \& Kirshner \cite{rpk96b}
found indications that the Galactic ratios between selective absorption and
color excess are similar for host galaxies in the nearby ($z \leq 0.1$) Hubble
flow.  Yet, what if these ratios changed with lookback time (e.g.,
\cite{agu99a})?  Could an evolution in dust grain size descending from
ancestral interstellar ``pebbles'' at higher redshifts cause us to
underestimate the extinction?  Large dust grains would not imprint the
reddening signature of typical interstellar extinction upon which our
corrections rely.

   However, viewing our SNe through such gray interstellar grains would also
induce a {\it dispersion} in the derived distances. Using the results of
Hatano, Branch, \& Deaton \cite{hat98}, Riess {\it et al.}  \cite{rie98b} estimate
that the expected dispersion would be 0.40 mag if the mean gray extinction were
0.25 mag (the value required to explain the measured MLCS distances without a
cosmological constant).  This is significantly larger than the 0.21 mag
dispersion observed in the high-redshift MLCS distances.  Furthermore, most of
the observed scatter is already consistent with the estimated {\it statistical}
errors, leaving little to be caused by gray extinction.  Nevertheless, if we
assumed that {\it all} of the observed scatter were due to gray extinction, the
mean shift in the SN~Ia distances would only be 0.05 mag.  With the
observations presented here, we cannot rule out this modest amount of gray
interstellar extinction.

  Gray {\it intergalactic} extinction could dim the SNe without either
telltale reddening or dispersion, if all lines of sight to a given
redshift had a similar column density of absorbing material.  The
component of the intergalactic medium with such uniform coverage
corresponds to the gas clouds producing Lyman-$\alpha$ forest
absorption at low redshifts.  These clouds have individual H~I column
densities less than about $10^{15} \, {\rm cm^{-2}}$ \cite{bah96}.
However, they display low metallicities, typically less than
10\% of solar. Gray extinction would require larger dust grains which
would need a larger mass in heavy elements than typical interstellar
grain size distributions to achieve a given extinction. It is possible
that large dust grains are blown out of galaxies by radiation pressure, 
and are therefore not associated with Lyman-$\alpha$ clouds \cite{agu99b}.

\bigskip

\hbox{
\hskip -0.45truein
\vbox{\hsize 3.5 truein
\psfig{figure=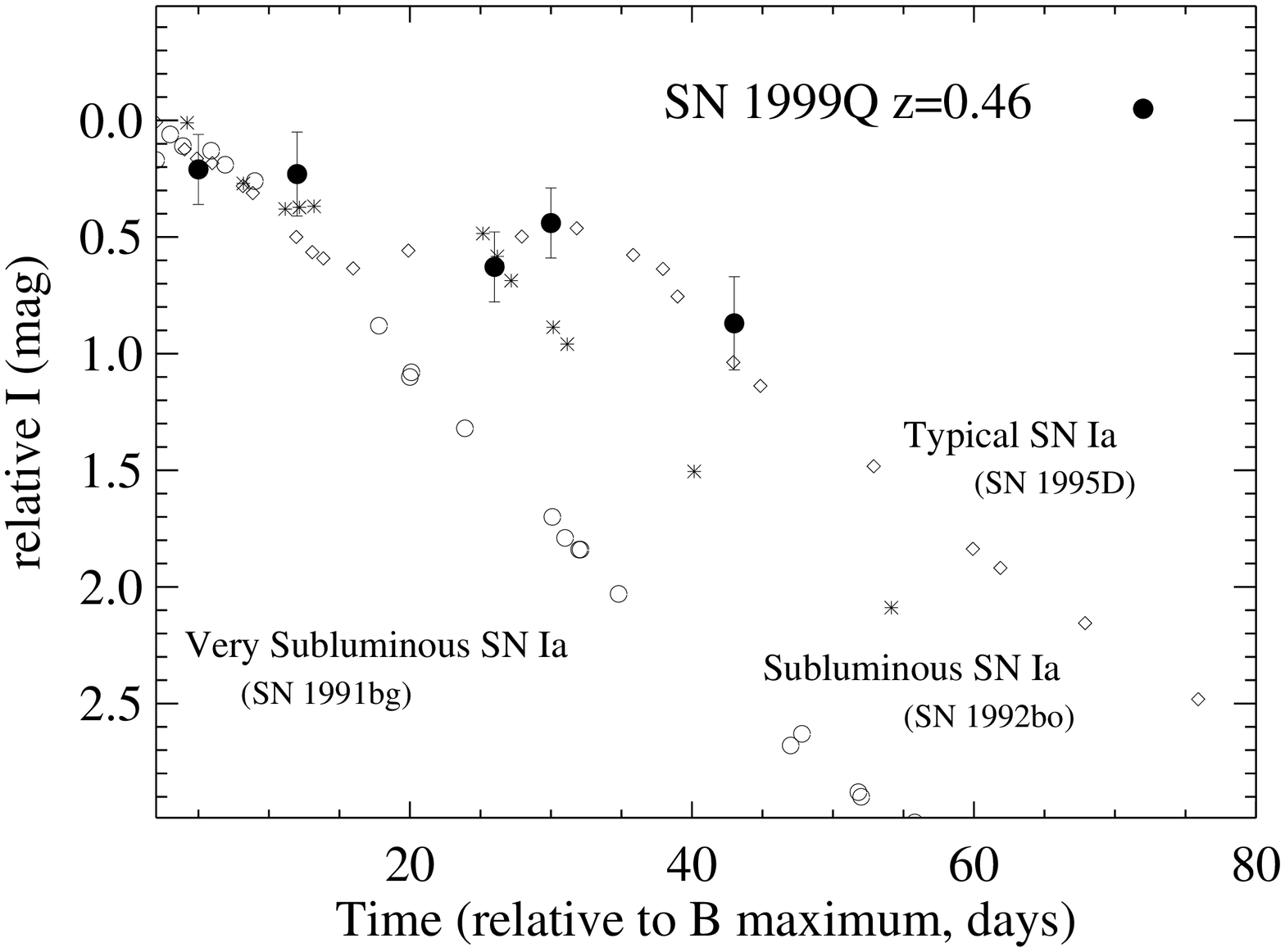,height=3.5truein,angle=90}
}
\hskip -0.0truein
\vbox{\hsize 3.0 truein
\psfig{figure=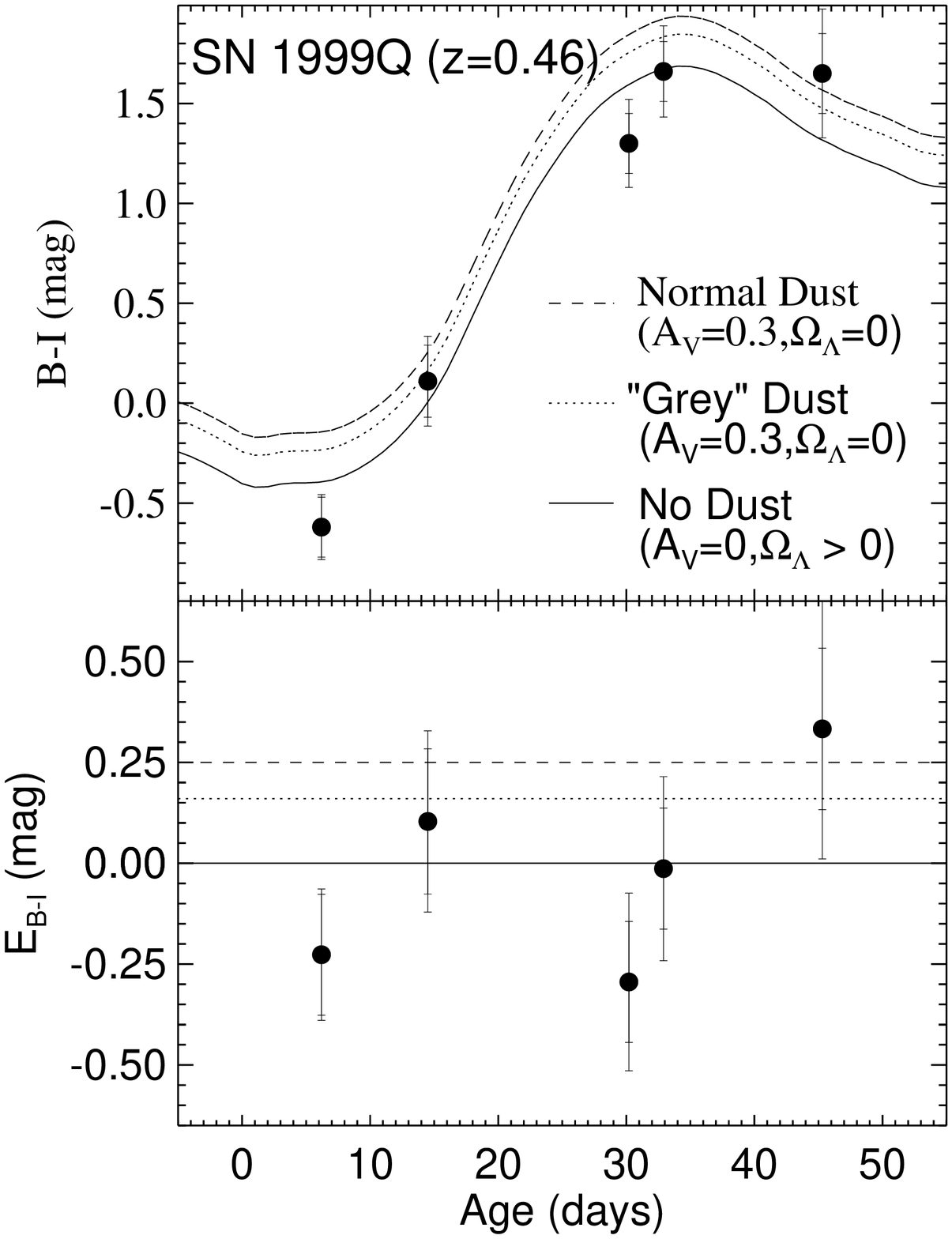,height=3.0truein,angle=0}
}
}

\smallskip

\noindent
{\it Figure 9 (left):} Restframe $I$-band (observed $J$-band) light curve of SN
1999Q ($z = 0.46$, 5 solid points; \cite{rie00a}), superposed on the
$I$-band light curves of several nearby SNe~Ia. Subluminous SNe~Ia exhibit 
a less prominent second maximum than do normal SNe~Ia.

\noindent
{\it Figure 10 (right):} Color excess, $E_{B-I}$, for SN 1999Q
and different dust models \cite{rie00a}. The data are most consistent
with no dust and $\Omega_\Lambda > 0$.

\bigskip

  But even the dust postulated by Aguirre \cite{agu99a,agu99b,agu99c} is not
{\it completely} gray, having a size of about 0.1~$\mu$m. We can test for such
nearly gray dust by observing high-redshift SNe~Ia over a wide wavelength range
to measure the color excess it would introduce. If $A_V = 0.25$ mag, then
$E(U-I)$ and $E(B-I)$ should be 0.12--0.16 mag \cite{agu99a,agu99b}. If, on the
other hand, the 0.25 mag faintness is due to $\Lambda$, then no such reddening
should be seen.  This effect is measurable using proven techniques; so far,
with just one SN~Ia (SN 1999Q; Figure 10), our results favor the no-dust
hypothesis to better than 2$\sigma$ \cite{rie00a}, but more work along
these lines is certainly warranted.

   Suppose, though, that for some reason the dust is {\it very} gray, or our
color measurements are not sufficiently precise to rule out Aguirre's (or
other) dust.  If the cumulative amount of gray dust along the line of sight
grows linearly with increasing redshift, we expect that the deviation of the
SN~Ia peak apparent magnitude from the low-$\Omega_M$, zero-$\Lambda$ model
(Figure 3) will continue growing, to first order (Figure 11). If, on the other hand,
the observed faintness of high-$z$ SNe~Ia is a consequence of positive
$\Lambda$, the deviation should actually begin to {\it decrease} at $z \approx
0.8$ (Figure 11). In essence, we are looking so far back in time that the
$\Lambda$ effect becomes small compared with $\Omega_M$; the Universe is
decelerating at that epoch. Thus, we are embarking on a campaign to find and
monitor $z = 0.8$--1.2 SNe~Ia. Given the expected uncertainties (Figure 11), a
sample of 10--20 SNe~Ia should give a good statistical result.

  Note that this test also applies to other systematic effects that grow
monotonically with redshift, as may be expected of possible evolution of the
white dwarf progenitors (e.g., \cite{hof98,ume99}), or gravitational lensing
\cite{wam98}. Indeed, this is our most decisive test to distinguish between
$\Lambda$ and systematic effects. Unless evolution of dust, or of the
progenitors, or of the lenses is fixed in such a way as to mimic the effects of
$\Lambda$ (e.g., \cite{dre00}), our claim of $\Omega_\Lambda > 0$ will become
much more convincing if the deviation of apparent magnitude decreases in the
expected manner. Such a turnover (Figure 11) can be considered the ``smoking
gun'' of $\Lambda$.

\medskip

\hbox{
\hskip +.6truein
\vbox{\hsize 4.0 truein
\psfig{figure=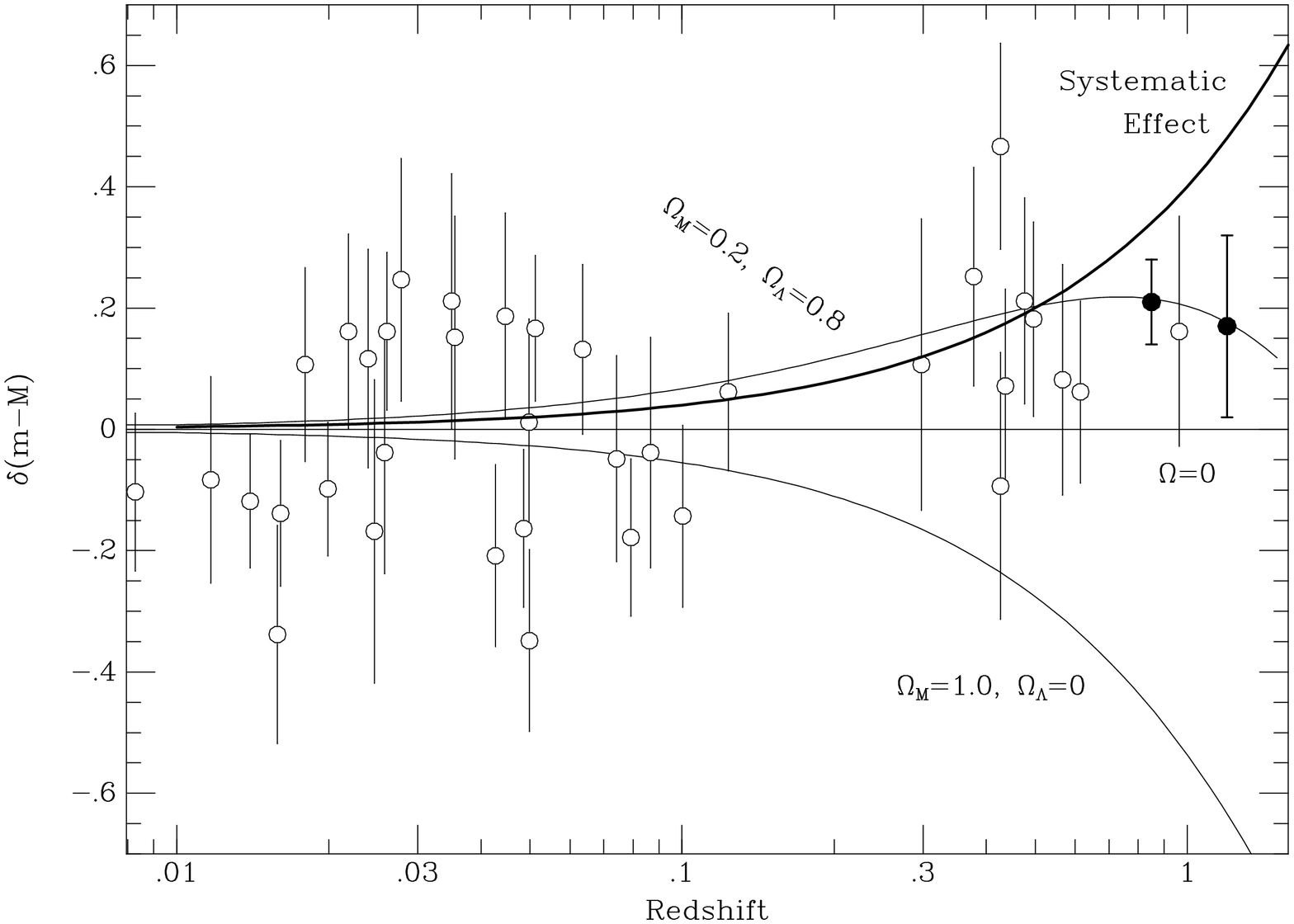,height=4.0truein,angle=270}
}
}

\medskip
\noindent
{\it Figure 11:} The HZT SN~Ia data from Figure 3 {\it (open circles)} are plotted 
relative to an empty universe {\it (horizontal line)}.  The two faint curves 
are the best-fitting $\Lambda$ model,
and the $\Omega_M = 1$ ($\Omega_\Lambda = 0$) model. The darker curve shows a
systematic bias that increases linearly with $z$ and is consistent with our $z
= 0.5$ data. The expected observational uncertainties of hypothetical 
SNe~Ia at redshifts of 0.85 and 1.2 are shown {\it (filled circles)}.

\bigskip

\section*{ CONCLUSIONS}

   The luminosity distances of the high-redshift Type Ia supernovae studied by
the High-z Supernova Search Team exceed the prediction of a low mass-density
($\Omega_M$ $\approx 0.2$) universe by about 0.25 mag.  A cosmological
explanation is provided by a positive cosmological constant at roughly the
3$\sigma$ confidence level, with the prior belief that $\Omega_M \geq 0$. We
also find that the expansion of the Universe is currently accelerating ($q_0
\leq 0$, where $q_0 \equiv \Omega_M/2 - \Omega_\Lambda$). The independent
results of the Supernova Cosmology Project are fully consistent with these
conclusions. Moreover, recent precise measurements of the cosmic microwave
background radiation strongly suggest that the Universe is flat ($\Omega_{\rm
total} = \Omega_M + \Omega_\Lambda = 1$); hence, if $\Omega_M \approx 0.3$ (as
suggested by many studies, such as of clusters of galaxies), then about 70\% of
the energy density of the Universe must consist of vacuum energy whose precise
nature and evolution are unknown (but definitely not radiation, normal matter,
or dark matter). Using the best current values of the Hubble constant,
$\Omega_M$, and $\Omega_\Lambda$, we find that the dynamical age of the
Universe is 14.2 $\pm 1.7$ Gyr, including systematic uncertainties in the
Cepheid distance scale used for the host galaxies of three nearby SNe~Ia. This
value is comparable to the derived ages of globular star clusters.

    Though the consistent results from the microwave background experiments are
reassuring, we are in the process of testing as exhaustively as possible all
systematic biases that could be affecting the SN~Ia results. For example,
qualitative comparisons of spectra of low-$z$ and high-$z$ SNe~Ia do not reveal
obvious differences, and quantitative tests are in progress. Moreover, the
restframe $I$-band light curves of low-$z$ SNe~Ia and a single measured
high-$z$ SN~Ia look similar, as do their broadband colors. The risetimes of
low-$z$ and high-$z$ SNe~Ia may differ a little, but the statistical
significance of this result is not high, and in any case the early part of the
light curve may have little bearing on the peak luminosity. Further tests are
in progress. Compelling evidence for acceleration may come in the next few
years from a comparison of the peak apparent brightness of $z \gtrsim 0.8$
SNe~Ia with the predictions of various models; the signature of nonzero
$\Lambda$ is quite distinct from that of dust, SN evolution, or other effects
that grown with redshift.

\section*{ ACKNOWLEDGMENTS}

  We thank all of our collaborators in the HZT for their contributions to this
work. A.V.F.'s supernova research at U.C. Berkeley is supported by NSF grants
AST-9417213 and AST-9987438, and by grants GO-7505 and GO-8177 from the Space
Telescope Science Institute, which is operated by the Association of
Universities for Research in Astronomy, Inc., under NASA contract
NAS~5-26555. A.V.F. is grateful to the meeting organizers for travel funds.


\begin{references}

\bibitem{avf97b}Filippenko, A. V., {\it ARAA}, {\bf 35}, 309 (1997b).
\bibitem{bra92}Branch, D., \& Tammann, G. A., {\it ARAA}, {\bf 30}, 359 (1992).
\bibitem{bra98}Branch, D., {\it ARAA}, {\bf 36}, 17 (1998).
\bibitem{bra93}Branch, D., \& Miller, D. L., {\it ApJ}, {\bf 405}, L5 (1993).
\bibitem{rie97}Riess, A. G., {\it et al.}, {\it AJ}, {\bf 114}, 722 (1997).
\bibitem{bfn93}Branch, D., Fisher, A., \& Nugent, P., {\it AJ}, {\bf 106}, 2383
  (1993).
\bibitem{vau95}Vaughan, T. E., Branch, D., Miller, D. L., \& Perlmutter, S., 
  {\it ApJ}, {\bf 439}, 558 (1995).
\bibitem{san96}Sandage, A., {\it et al.}, {\it ApJ}, {\bf 460}, L15 (1996).
\bibitem{sah97}Saha, A., {\it et al.}, {\it ApJ}, {\bf 486}, 1 (1997).
\bibitem{avf97a}Filippenko, A. V., in {\it Thermonuclear Supernovae}, 
  ed. P. Ruiz-Lapuente, {\it et al.} (Dordrecht: Kluwer), p. 1 (1997a).
\bibitem{van92}van den Bergh, S., \& Pazder, J., {\it ApJ}, {\bf 390}, 34 (1992).
\bibitem{san93}Sandage, A., \& Tammann, G. A., {\it ApJ}, {\bf 415}, 1 (1993).
\bibitem{avf92b}Filippenko, A. V., {\it et al.}, {\it ApJ}, {\bf 384}, L15 (1992b).
\bibitem{phi92}Phillips, M. M., {\it et al.}, {\it AJ}, {\bf 103}, 1632 (1992).
\bibitem{avf92a}Filippenko, A. V., {\it et al.}, {\it AJ}, {\bf 104}, 1543 (1992a).
\bibitem{lei93}Leibundgut, B., {\it et al.}, {\it AJ}, {\bf 105}, 301 (1993).
\bibitem{tur96}Turatto, M., {\it et al.}, {\it MNRAS}, {\bf 283}, 1 (1996).
\bibitem{sun96a}Suntzeff, N., in {\it Supernovae and Supernova Remnants}, 
  ed. R. McCray \& Z. Wang (Cambridge: Cambridge Univ. Press), p. 41 (1996).
\bibitem{wli00a}Li, W. D., {\it et al.}, in {\it Cosmic Explosions}, ed. S. S.
  Holt \& W. W. Zhang (New York: American Inst. Physics), p. 91 (2000a).
\bibitem{psk77}Pskovskii, Yu. P., {\it Sov. Astron.}, {\bf 21}, 675 (1977).
\bibitem{psk84}Pskovskii, Yu. P., {\it Sov. Astron.}, {\bf 28}, 658 (1984).
\bibitem{bra81}Branch, D., {\it ApJ}, {\bf 248}, 1076 (1981).
\bibitem{phi93}Phillips, M. M., {\it ApJ}, {\bf 413}, L105 (1993).
\bibitem{ham95}Hamuy, M., {\it et al.}, {\it AJ}, {\bf 109}, 1 (1995).
\bibitem{ham96b}Hamuy, M., {\it et al.}, {\it AJ}, {\bf 112}, 2398 (1996b).
\bibitem{tri97}Tripp, R., {\it A\&A}, {\bf 325}, 871 (1997).
\bibitem{rpk95}Riess, A. G., Press, W. H., \& Kirshner, R. P., 
   {\it ApJ}, {\bf 438}, L17 (1995).
\bibitem{rpk96a}Riess, A. G., Press, W. H., \& Kirshner, R. P., 
   {\it ApJ}, {\bf 473}, 88 (1996a).
\bibitem{tri98}Tripp, R., {\it A\&A}, {\bf 331}, 815 (1998).
\bibitem{rie00b}Riess, A. G., {\it et al.}, in preparation (2000b).
\bibitem{ham96a}Hamuy, M., {\it et al.}, {\it AJ}, {\bf 112}, 2391 (1996a).
\bibitem{bra96}Branch, D., Romanishin, W., \& Baron, E., {\it ApJ}, 
   {\bf 465}, 73; erratum {\bf 467}, 473 (1996).
\bibitem{wli00b}Li, W. D., {\it et al.}, in {\it Cosmic Explosions}, ed. S. S.
  Holt \& W. W. Zhang (New York: American Inst. Physics), p. 103 (2000b).
\bibitem{avf00a}Filippenko, A. V., {\it et al.}, in preparation (2000a).
\bibitem{goo95}Goobar, A., \& Perlmutter, S., {\it ApJ}, {\bf 450}, 14
  (1995).
\bibitem{nor89}Norgaard-Nielsen, H., {\it et al.}, {\it Nature}, 
   {\bf 339}, 523 (1989).
\bibitem{per97}Perlmutter, S., {\it et al.}, {\it ApJ}, {\bf 483}, 565 (1997).
\bibitem{sch98}Schmidt, B. P., {\it et al.},  {\it ApJ}, {\bf 507}, 46 (1998).
\bibitem{per95}Perlmutter, S., {\it et al.}, {\it IAUC} 6270 (1995).
\bibitem{sun96b}Suntzeff, N., {\it et al.}, {\it IAUC} 6490 (1996b).
\bibitem{kim96}Kim, A., Goobar, A., \& Perlmutter, S., {\it PASP}, 
   {\bf 108}, 190 (1996).
\bibitem{lei96}Leibundgut, B., {\it et al.}, {\it ApJ}, {\bf 466}, L21 (1996).
\bibitem{gol97}Goldhaber, G., {\it et al.}, in {\it Thermonuclear Supernovae}, 
  ed. P. Ruiz-Lapuente, {\it et al.} (Dordrecht: Kluwer), p. 777 (1997).
\bibitem{avf00b}Filippenko, A. V., {\it et al.}, in preparation (2000b).
\bibitem{gar98a}Garnavich, P., {\it et al.}, {\it ApJ}, {\bf 493}, L53 (1998a).
\bibitem{rie98a}Riess, A. G., Nugent, P. E., Filippenko, A. V., Kirshner,
  R. P., \& Perlmutter, S., {\it ApJ}, {\bf 504}, 935 (1998a).
\bibitem{avf98}Filippenko, A. V., \& Riess, A. G., {\it Physics 
   Reports}, {\bf 307}, 31 (1998).
\bibitem{rie98b}Riess, A. G., {\it et al.}, {\it AJ}, {\bf 116}, 1009 (1998b).
\bibitem{per99}Perlmutter, S., {\it et al.}, {\it ApJ}, {\bf 517}, 565 (1999).
\bibitem{zal97}Zaldarriaga, M., Spergel, D. N., \& Seljak, U.
     {\it ApJ}, {\bf 488}, 1 (1997).
\bibitem{eis98}Eisenstein, D. J., Hu, W., \& Tegmark, M. {\it ApJ}, {\bf
     504}, L57 (1998).
\bibitem{han98}Hancock, S., Rocha, G., Lazenby, A. N., \& Guti\'{e}rrez, C. M.,
    {\it MNRAS}, {\bf 294}, L1 (1998).
\bibitem{lin98}Lineweaver, C. H., \& Barbosa, D., {\it ApJ}, {\bf 496}, 
   624 (1998).
\bibitem{gar98b}Garnavich, P., {\it et al.}, {\it ApJ}, {\bf 509}, 74 (1998b).
\bibitem{lin98}Lineweaver, C. H., {\it ApJ}, {\bf 505}, L69 (1998).
\bibitem{efs99}Efstathiou, G., {\it et al.}, {\it MNRAS}, {\bf 303}, L47 (1999).
\bibitem{deb00}de Bernardis, P., {\it et al.}, {\it Nature}, {\bf 404}, 955 (2000).
\bibitem{han00}Hanany, S., {\it et al.}, {\it ApJ}, submitted, astro-ph/0005123 (2000).
\bibitem{bal00}Balbi, A., {\it et al.}, {\it ApJ}, submitted, astro-ph/0005124 (2000).
\bibitem{mil99}Miller, A. D., {\it et al.}, {\it ApJ}, {\bf 524}, 1 (1999).
\bibitem{bah99}Bahcall, N. A., Ostriker, J. P., Perlmutter, S., \&
  Steinhardt, P. J., {\it Science}, {\bf 284}, 1481 (1999).
\bibitem{osw96}Oswalt, T. D., Smith, J. A., Wood, M. A., \& Hintzen, P.,
   {\it Nature}, {\bf 382}, 692 (1996).
\bibitem{cow97}Cowan, J. J., McWilliam, A., Sneden, C., \& Burris, D. L.,
   {\it ApJ}, {\bf 480}, 246 (1997).
\bibitem{gra97}Gratton, R. G., Fusi Pecci, F., Carretta, E., Clementini, G.,
  Corsi, C. E., \& Lattanzi, M., {\it ApJ}, {\bf 491}, 749 (1997).
\bibitem{cha98}Chaboyer, B., Demarque, P., Kernan, P. J., \& Krauss, L. M.,
   {\it ApJ}, {\bf 494}, 96 (1998).
\bibitem{hof98}H\"{o}flich, P., Wheeler, J. C., \& Thielemann, F. K.,
  {\it ApJ}, {\bf 495}, 617 (1998).
\bibitem{ume99}Umeda, H., {\it et al.}, {\it ApJ}, {\bf 522}, L43 (1999).
\bibitem{dom99}Dom\'\i nguez, I., H\"{o}flich, P., Straniero, O., \& Wheeler, J.,
  in {\it Future Directions of Supernovae Research: Progenitors
  to Remnants} (Assergi, Italy), astro-ph/9905047 (1999).
\bibitem{yun00}Yungelson, L. R., \& Livio, M., {\bf ApJ}, {\bf 528}, 108
  (2000).
\bibitem{nom00} Nomoto, K., Umeda, H., Hachisu, I., Kato, M., Kobayashi, C., \&
   Tsujimoto, T., in {\it Type Ia Supernovae: Theory and Cosmology},
    ed. J. Truran \& J. Niemeyer (Cambridge: Cambridge Univ. Press), 
   in press, astro-ph/9907386 (2000).
\bibitem{dre00}Drell, P. S., Loredo, T. J., \& Wasserman, I., {\it ApJ},
   {\bf 530}, 593 (2000).
\bibitem{cap97}Cappellaro, E., {\it et al.}, {\it A\&A}, {\bf 322}, 431 (1997).
\bibitem{nug95}Nugent, P., Phillips, M., Baron, E., Branch, D., \&
   Hauschildt, P., {\it ApJ}, {\bf 455}, L147 (1995).
\bibitem{coi00}Coil, A. L., {\it et al.}, submitted (2000).
\bibitem{for93}Ford, C. H., {\it et al.}, {\it AJ}, {\bf 106}, 1101 (1993).
\bibitem{rie00a}Riess, A. G., {\it et al.}, {\it ApJ}, {\bf 536}, 62 (2000).
\bibitem{rie99b}Riess, A. G., et al., {\it AJ}, {\bf 118}, 2675 (1999b).
\bibitem{rie99a}Riess, A. G., Filippenko, A. V., Li, W. D., \& Schmidt,
  B. P., {\it AJ}, {\bf 118}, 2668 (1999a).
\bibitem{gol98a}Goldhaber, G., {\it et al.}, {\it BAAS}, {\bf 30}, 1325 (1998a).
\bibitem{gol98b}Goldhaber, G., {\it et al.}, in {\it Gravity: From the Hubble Length
  to the Planck Length}, SLAC Summer Institute (Stanford, CA: SLAC) (1998b).
\bibitem{gro98}Groom, D. E., {\it BAAS}, {\bf 30}, 1419 (1998).
\bibitem{ald00}Aldering, G., Knop, R., \& Nugent, P. {\it AJ}, {\bf 119}, 2110
  (2000).
\bibitem{rpk96b}Riess, A. G., Press, W. H., \& Kirshner, R. P.,
   {\it ApJ}, {\bf 473}, 588 (1996b).
\bibitem{agu99a}Aguirre, A. N., {\it ApJ}, {\bf 512}, L19 (1999a).
\bibitem{hat98}Hatano, K., Branch, D., \& Deaton, J., {\it ApJ},
   {\bf 502}, 177 (1998).
\bibitem{bah96}Bahcall, J. N., {\it et al.}, {\it ApJ}, {\bf 457}, 19 (1996).
\bibitem{agu99b}Aguirre, A. N., {\it ApJ}, {\bf 525}, 583 (1999b).
\bibitem{agu99c}Aguirre, A., \& Haimin, Z., {\it ApJ}, {\bf 525}, 583 
  (1999).
\bibitem{wam98}Wambsganss, J., Cen, R., \& Ostriker, J. P.,   
   {\it ApJ}, {\bf 494}, 29 (1998).

\end{references}
\end{document}